\documentclass[12pt]{article}

\usepackage{hyperref}

\usepackage{epsf}
\usepackage{amsmath,amssymb}
\usepackage{latexsym}
\usepackage{graphicx,color}
\usepackage{wrapfig}
\usepackage{latexsym}
\usepackage{graphics}

\usepackage{color}
\usepackage{delarray,color}
\usepackage{fancybox}  
\newcommand {\beq}{\begin{align}}
\newcommand {\eeq}{\end{align}}

\newcommand{\be}{\begin{equation}}
\newcommand{\ba}{\begin{align}}
\newcommand{\ea}{\end{align}}
\newcommand{\ee}{\end{equation}}

\newcommand{\beqa}{\begin{align}}
\newcommand{\eeqa}{\end{align}}
\newcommand{\CR}{\nonumber \\}

\newcommand{\unit}{\hbox to 3.8pt{\hskip1.3pt \vrule height 7.4pt
    width .4pt \hskip.7pt \vrule height 7.85pt width .4pt \kern-2.4pt
    \hrulefill \kern-3pt \raise 3.7pt\hbox{\char'40}}}

\def\matt[#1,#2,#3,#4]{\left(%
\begin{array}{cc} #1 & #2 \\ #3 & #4 \end{array} \right)}

\newcommand{\ket}[1]{{\left| #1 \right\rangle}}
\newcommand{\bra}[1]{{\left\langle#1\right|}}
\usepackage{tabularx}

\catcode`\@=11
\@addtoreset{equation}{section}

\catcode`@=12
\relax

\begin{document}
%%%%%%%%%%%%%%%%%%%%%%%%%%%%%%%%%%%%%%%%%%%%%%%%%%%%%%%%%%%%%%%%%%%%%%%%
%\baselineskip 0.7cm

\begin{titlepage}

%% Set the number of the title with 0
\setcounter{page}{0}

%% change the footnote symbol
\renewcommand{\thefootnote}{\fnsymbol{footnote}}

\begin{flushright}
%{\tt 
YITP-20-60
%\\}
\end{flushright}

\vskip 1.35cm

\begin{center}
{\Large \bf 
Bulk Locality in AdS/CFT Correspondence
}

\vskip 1.2cm 

{\normalsize
Seiji Terashima\footnote{terasima(at)yukawa.kyoto-u.ac.jp}
}

\vskip 0.8cm

{ \it
Yukawa Institute for Theoretical Physics, Kyoto University, Kyoto 606-8502, Japan
}

\end{center}

\vspace{12mm}

\centerline{{\bf Abstract}}

In this paper, 
we study the bulk local states in AdS/CFT correspondence
in the large $N$ limit using the formula explicitly relating 
the bulk local operators and the CFT local operators.
We identify the bulk local state in terms of CFT local states and 
find that the bulk local state corresponds to  
a CFT state supported in the whole space, which means a version of the subregion duality is not valid.
On the other hand,
CFT states supported in a space region are expressed
in terms of the bulk states supported in a certain region.
We find that the quantum error correction proposal is
not realized although 
the puzzles of the radial locality which motivated the proposal
are resolved in our analysis.
For the understating of the bulk local states,
an explicit realization of an analogue of the Reeh-Schlieder theorem
plays an important role.

\end{titlepage}
\newpage

\tableofcontents
\vskip 1.2cm 

\section{Introduction and summary}

The $AdS/CFT$ correspondence \cite{Maldacena} is highly non-trivial and
important in various aspects of physics and has been investigated intensively.
There is no proof
for this correspondence, thus it is still a conjecture
although there are many evidences.

The key ingredient of the $AdS/CFT$ correspondence is the locality in the bulk, which is needed for the emergence of the extra dimension (for a recent discussion on the bulk locality, see \cite{Giddings, Banerjee:2016mhh}).
In particular, the local states in the bulk should be understood from the states of $CFT_d$.
In order to study the correspondence between the bulk theory 
and the CFT, some  bulk-boundary correspondence, like the GKPW relation \cite{GKP, W} or the BDHM relation \cite{BDHM} is usually assumed. 
Instead of assuming such bulk-boundary correspondences, 
with the natural assumptions on the large $N$ strong couping $CFT_d$,
we can show the equivalence between the spectrum (i.e. the states and the Hamiltonian)
of the $CFT_d$ and the spectrum of the bulk theory on $AdS_{d+1}$ in the (naive) large $N$ limit, in which 
the bulk gravity theory will become a free theory \cite{op}.\footnote{ 
Here,
we considered the (asymptotically) global $AdS_{d+1}$ only, because the 
spectrum of the theory will be continuous and ambiguous for the Poincare patch of AdS.} 
Here, the Hamiltonian is given by the dilatation operator. 
Then, the bulk local operator and the bulk local state are explicitly given by the $CFT_d$ operator and the $CFT_d$ state, using
the explicit identification of the energy eigen states in $CFT_d$ 
to the those in the bulk theory on $AdS_{d+1}$.\footnote{
A local state in this paper means a state obtained by acting a local operator on the vacuum.
More precisely, the local operators and states should be smeared over a region in the spacetime.}
Furthermore, it was shown that the GKPW and BDHM relations are consequences of this identification.

Note that these results in \cite{op}  are based on some earlier works 
\cite{BKL, BDHM, Pol} (see also \cite{FK}-\cite{Rehren}) and
the identification in \cite{op}
may be consistent with the HKLL reconstruction \cite{HKLL}
in which assuming the BDHM relation \cite{BDHM} which relates 
the $CFT_d$ primary field as a limit of bulk field to the boundary,
the bulk local operator is reconstructed from the $CFT_d$ operator.
%Because this is for free field in the bulk,
%the bulk local state is also reconstructed from the CFT state, 
%by just acting the field on the vacuum.
Moreover, the bulk local state at the center of a time slice of $AdS_{d+1}$ is also explicitly reconstructed by
imposing the invariance under some ``rotations'' symmetries of $AdS_{d+1}$ which are expected to fix the center \cite{Ta,NO1,Ver}.
The results in \cite{op} are consistent with these also.\footnote{
For a state at the center, we can check that the construction in \cite{op} gives same state given by \cite{Ta,NO1,Ver}.
However, there are states which are invariant under the ``rotations'' symmetries , but is not a local state \cite{op}. In \cite{Goto2}, the bulk local states not at the center are constructed using the conformal transformation also for $CFT_2$. These could be different from the ones we will use
in this paper because the local operator with spacetime derivatives is a local operator.}

In \cite{op}, 
the bulk locality in the large $N$ limit comes from 
the large $N$ factorization, which guarantees the theory is (generalized) free
and
the completeness condition of the spectrum, which gives the Fock space
together with the conformal symmetry, which may guarantee the Lorentz symmetry
for the local neighborhood of a spacetime point.
On the other hand, 
in \cite{ADH} 
some problems or puzzles related to the locality in the radial direction in the bulk
were given %, following \cite{}, 
and the quantum error correction codes are claimed to be the key to
solve these problems (for a  review, see \cite{Ha}).
However, the structure of the quantum  error correction codes has not been explicitly identified 
in realistic models of the $AdS/CFT$ correspondence.
Thus, it is worth to investigate these puzzles 
in the bulk local states or the bulk local operator expressed by 
the energy eigen states of $CFT_d$ in the large $N$ limit 
using the results in \cite{op}.

In this paper, 
we show how these puzzles of the radial locality are solved 
in terms of the bulk and CFT local operators or states which are given by the energy eigen states of $CFT_d$.\footnote{
In this paper, we only consider 
the bulk (free) scalar field and 
the corresponding sector of CFT in the naive large $N$ limit, for simplicity. 
}
The resolution is partly based on the fact that
we need the ``cut-off'' for the high energy states in the low energy effective theory
which is regarded as the bulk (gravity) theory.
Note that this cut-off is crucial for the error correction code proposal \cite{ADH}. Indeed, it is about low energy states (or code subspace), 
thus it is obvious that the cut-off is needed, even if we consider the large N limit.
In other words, in the formal large $N$ expansion, we can not resolve the puzzles.
We need the large, but finite $N$ effects (i.e. finite plank length effects) for the resolution of them.

With this cut-off,
we explicitly identify the bulk local state in terms of CFT local states.
We also explicitly identify CFT states supported in a space region 
in terms of the bulk local states.
We find that the bulk local state corresponds to  
a CFT state supported in the whole space $S^{d-1}$, not a subregion.
This means that the subregion duality \cite{Wa}-\cite{Bo} is not valid in our model.\footnote{In this paper, the subregion duality means that the operators supported in the CFT subregion $A$ is equivalent to the operators supported in a corresponding bulk subregion $a$. Another weaker version of it is that the CFT density matrix for the region $A$, which is defined by the tracing out, is equivalent to the density matrix for the bulk theory defined in a bulk region $a$ with an appropriate boundary condition. This will be consistent with our results.
}
On the other hand, we argue that a CFT state supported in a space region $A$ corresponds to a  
a bulk state supported in a certain space region.
We call this bulk region as the minimal surface wedge of $A$, which is a kind of generalization of 
the entanglement wedge.
Furthermore, we find that two bulk local states at a same spacetime point constructed from CFT states supported in space regions $A$
and $A'$ are different generically if $A \neq A'$, even in the low energy theory.
This means that the quantum error correction code proposal \cite{ADH} is 
not realized % in our model.
although the puzzles of the radial locality raised in \cite{ADH} are resolved.

One of the important topics in this paper 
is the Reeh-Schlieder theorem.
We explicitly construct a CFT local state at a spacetime point which is
effectively equivalent to a CFT local state at a different spacetime point.
Hence, it is important to find which state is effectively local with the cut-off we introduced
and then
we call them the CFT effectively local states at each spacetime point.
The CFT effectively local states on the time slice $t=0$ do not span even the low energy states.
However, by the time-evolution, a CFT effectively local state at $t = t_0 \neq 0$ gives 
a CFT state supported in a region at $t=0$, which is not given by effectively local states on $t=0$.
If we consider the CFT effectively local states in whole spacetime (or the spacetime region $\pi \leq t \leq \pi$) and consider 
the time evolution of them to the time slice $t=0$,
then any low energy state is expected to be given by them
because in the low energy theory the effectively local states play the role of the local states.
\footnote{
Note that any CFT state is given by a superposition of the CFT local states in the $t=0$ time slice 
of spacetime. For a small number of (free) fields, any state (without a cut-off)
will be given by a effectively local states in the time slice, which implies that there is no bulk dual for such CFT.
}
We will see that
the maximum distance 
from the boundary in radial direction $z$
of the bulk local state at $t=0$ constructed from 
such a CFT effectively local state at $t=t_0$
is $|t_0|$ because of the bulk causality.

In this paper, we concentrate on the free field limit of the bulk theory around the AdS spacetime.
A generalization of \cite{op} to general classical backgrounds
was done in \cite{classical} and showed that the Einstein equation was derived with assumptions made in \cite{op}.
It is interesting to investigate how the analysis in the paper can be generalized to a general back ground.
In particular, for the black hole, we expect that the brick wall \cite{tHooft} will appear \cite{brick} instead of the horizon
and it is interesting to study how the locality near the brick wall is modified.

This paper is organized as follows.
In the next section
we review the relation between the free scalar field on $AdS_{d+1}$ and the large $N$ $CFT_d$ according to \cite{op}.
Based on this, we will describe the bulk local states and 
the CFT local states in terms of the energy eigen states.
In section three, 
we study the bulk locality in AdS/CFT correspondence
in the naive large $N$ limit.
We explicitly identify the bulk local state as a CFT state
and we also identify a CFT state supported in a space region 
as a bulk state.

%In the appendices, 

\section{Bulk local states %AdS_{d+1}$ in the $CFT_d$ 
 and CFT local states
}

In this section,
we will review the relation between free scalar field on $AdS_{d+1}$ and large $N$ $CFT_d$ according to \cite{op}.
Based on this, we will describe the bulk local states and 
the CFT local states in terms of the energy eigen states.

\subsection{Relation between free scalar field on $AdS_{d+1}$ and large $N$ $CFT_d$}

In this subsection,
we will review local operators and states of the (free) scalar field on $AdS_{d+1}$
in terms of $CFT_d$ in the large $N$ limit, 
following \cite{op}.%, in particular, use the (almost) same notation.

Let us consider the free scalar field for which
the action is given by 
\begin{align} 
S_{scalar}= \int d^{d+1} x \sqrt{-\det (g)}
\left(
\frac12 g^{M N} \nabla_M \phi \nabla_N \phi
+ \frac{m^2}{2} \phi^2 
\right),
\end{align}
where $M,N=1, \cdots,d+1$,
on global $AdS_{d+1}$.
The $AdS_{d+1}$ metric is
\begin{align}
d s^2_{AdS} = -(1+r^2) dt^2 +\frac{1}{1+r^2} d r^2+ r^2 d \Omega_{d-1}^2,
\end{align}
where $0 \leq  r < \infty$, $-\infty < t < \infty$
and $d \Omega_{d-1}^2$ is the metric for the $d-1$-dimensional 
round unit sphere $S^{d-1}$.
We set the AdS scale $l_{AdS}=1$ in this paper.
By the coordinate change $r=\tan \rho$,
the metric is also written as
\begin{align}
d s^2_{AdS} = \frac{1}{\cos^2 (\rho)} 
\left( 
-dt^2 +d \rho^2+ \sin^2 (\rho)  d \Omega_{d-1}^2 
\right),
\end{align}
where 
$0 \leq  \rho < \pi/2$.
With $z=\pi/2- \rho$, the metric is given by 
\begin{align}
d s^2_{AdS} = \frac{1}{\sin^2 (z)} 
\left( 
-dt^2 +d z^2+ \cos^2 (z)  d \Omega_{d-1}^2 
\right).
\end{align}
The boundary of the $AdS_{d+1}$ is located at
$\rho=\pi/2$ or $z=0$.

We expand the quantized field $\hat{\phi}$ with the spherical harmonics $Y_{lm}(\Omega)$,
\begin{align}
\hat{\phi}(t,\rho,\Omega)=
\sum_{n,l,m} 
\left( 
\hat{a}_{nlm}^\dagger e^{i \omega_n t} 
+\hat{a}_{nlm} e^{-i \omega_n t} 
\right)
\psi_{nlm} (\rho)  Y_{lm} ( \Omega),
\end{align}
where $\Omega$ represents the coordinates of $S^{d-1}$ and
\begin{align}
\psi_{nl} (\rho)=\frac{1}{N_{nl}} \sin^l (\rho) \cos^\Delta (\rho)\,
{}_2 F_1 \left(
-n,\Delta+l+n,l+\frac{d}{2},\sin^2(\rho)
\right) ,
\end{align}
where $\Delta$ is given by $\Delta = d/2\pm \sqrt{m^2+d^2/4}$, and 
\begin{align}
\omega_{nl}=2n+l+\Delta,
\end{align}
where $n,l =0,1,2,3, \cdots$.
The normalization constant is given by
\begin{align}
N_{nl}=(-1)^n \sqrt{
\frac{n ! \Gamma(l+\frac{d}{2})^2 \Gamma(\Delta +n+1-\frac{d}{2})}
{\Gamma(n+l+\frac{d}{2}) \Gamma(\Delta+n+l)}
}.
\end{align}
The operators $\hat{a}_{nlm}$ satisfy
the commutation relation
\begin{align}
[\hat{a}_{nlm} , \hat{a}^\dagger_{n' l' m'}]=\delta_{n,n'} \delta_{l,l'} \delta_{m,m'},
\end{align}
and the Hamiltonian is such that
\begin{align}
[\hat{H}, \hat{a}_{n l m}]=-\omega_{nl}.
\end{align}
The Hilbert space is the Fock space 
spanned by $\prod_{n,l,m} (\hat{a}_{n l m}^\dagger)^{{\cal N}_{nlm}} | 0 \rangle$,
where ${\cal N}_{nlm}$ is a non-negative integer.
We choose the constant shift of the Hamiltonian as
$\hat{H} | 0 \rangle=0 $ where $ | 0 \rangle $ is the vacuum, i.e. $\hat{a}_{n l m} | 0 \rangle=0 $.

Now, let us consider a scalar primary field ${\cal O}_\Delta (x)$ 
in 
$CFT_d$
on $\mathbf{R} \times S^{d-1}$ where $\mathbf{R}$ is the time direction and 
the radius of $S^{d-1}$ is taken to be one.
For a review of the $CFT_d$, see for example, \cite{Qu, Ry,SD}.
Any state in CFT can be obtained as a linear combination of 
the primary state $| \Delta \rangle=\lim_{x \rightarrow 0} 
{\cal O}_\Delta (x) | 0 \rangle$ with the $\hat{P}^\mu$,
for example,
\begin{align}
\hat{P}^{ \mu_1} \hat{P}^{\mu_2} \cdots \hat{P}^{\mu_l } | \Delta \rangle.
\end{align}
We also define the operator 
\begin{align}
\hat{{\cal O}}_\Delta=\lim_{x \rightarrow 0} 
\hat{{\cal O}}^+_\Delta (x),
\end{align}
where
$\hat{{\cal O}}^+_\Delta (x)$
is the regular parts of ${\cal O}_\Delta (x)$ in $x^\mu \rightarrow 0$ limit\footnote{
This should be done after taking the large $N$ limit.
More precisely,  $\hat{{\cal O}}_\Delta$ is the sum of the operators 
of dimension $\Delta$ up to $1/N$ corrections in ${\cal O}_\Delta (x)$.
}
which can be expanded by the polynomial of $x^\mu$.\footnote{
For the stress energy tensor in two dimensional theory,
$\hat{{\cal O}}_\Delta$ is
 $L_{-2}$ or $\tilde{L}_{-2}$.}
This satisfies $| \Delta \rangle 
=\hat{{\cal O}}_\Delta | 0 \rangle$
where  $| 0 \rangle$ is the conformal vacuum.
It was shown in \cite{op} that 
the spectrum of this $CFT_d$ in the large $N$ limit is equivalent 
to the spectrum of free scalar on $AdS_{d+1}$
under some natural assumptions on the spectrum. 
The identification of the CFT states
to the states of the Fock space of  
the scalar fields in AdS
is explicitly given by the identification of the raising operators as
\begin{align}
\hat{a}_{n l m}^\dagger =
c_{nl} \, s^{\mu_1 \mu_2 \ldots \mu_l}_{(l,m)} P_{\mu_1} P_{\mu_2} \cdots P_{\mu_l} 
 (P^2)^n \hat{{\cal O}}_\Delta
\label{id1}
\end{align}
where $c_{nl}$ is the normalization constant,
which was determined in \cite{op},
$P^\mu$ act on an operator such that
$P^\mu \hat{\phi} =[\hat{P}^\mu, \hat{\phi}]$
and $s^{\mu_1 \mu_2 \ldots \mu_l}_{(l,m)}$ is
a normalized rank $l$ symmetric traceless constant tensor.

\subsection{Bulk local states }

We will decompose the local operator in the bulk description 
to positive and negative frequency modes as
\begin{align}
\hat{\phi}(t,\rho,\Omega) 
=
\hat{\phi}^+(t,\rho,\Omega)
+
\hat{\phi}^-(t,\rho,\Omega),
\end{align}
where $\hat{\phi}^-(t,\rho,\Omega) =(\hat{\phi}^+(t,\rho,\Omega))^\dagger$.
In this paper, we will concentrate on the one particle states in the naive large $N$ limit only.
For this, the commutator of the operators $\hat{\phi}(\rho,\Omega) \equiv \hat{\phi}(t=0,\rho,\Omega) $  
is a constant and given by the overlap of the corresponding states, $\hat{\phi}^+  (\rho,\Omega) \ket{0} $,
which is
\begin{align}
\bra{0} [ \hat{\phi}_1 (\rho,\Omega) , \hat{\phi}_2 (\rho',\Omega') ] \ket{0} 
&=
\bra{0} \hat{\phi}_1^- (\rho,\Omega) \,\, \hat{\phi}_2^+  (\rho',\Omega') \ket{0} 
-\bra{0}  \hat{\phi}_2^- (\rho',\Omega') \,\,  \hat{\phi}_1^+  (\rho,\Omega) \ket{0}
\CR
&=2 i \, {\rm Im} ( \bra{0} \hat{\phi}_1^- (\rho,\Omega) \,\, \hat{\phi}_2^+  (\rho',\Omega') \ket{0}),
\end{align}
where $\hat{\phi}_i$ represents  $\hat{\phi}$ with some space-time derivatives
and  $\hat{\phi}_i^\pm$ is 
the positive/negative frequency modes of $\hat{\phi}_i$.
Thus, we will study the states, instead of the operators
in order to understand the locality in the bulk.
In particular, the commutator of the operators is zero
if the corresponding overlap vanishes.

Using (\ref{id1}),
the bulk local operator is expressed as
\begin{align}
\hat{\phi}^+(t=0,\rho,\Omega)
&=\sum_{n,l,m}  
\psi_{nl} (\rho)  Y_{lm} ( \Omega)
\hat{a}_{nlm}^\dagger
\nonumber \\ 
&= \sum_{n,l,m}  
\psi_{nl} (\rho)  Y_{lm} ( \Omega)
c_{nl} s^{\mu_1 \mu_2 \ldots \mu_l}_{(l,m)} P_{\mu_1} P_{\mu_2} \cdots P_{\mu_l}  
(P^2)^n  \hat{{\cal O}}_\Delta,
\label{local}
\end{align}
where only the CFT operators appear 
in the last line.\footnote{
The derivatives of a local operator with $t,\rho, \Omega$ is also 
a local operator, for example, the momentum $\hat{\pi}(t,\rho,\Omega)=\sqrt{-g} g^{tt}
\frac{\partial }{\partial t}\hat{\phi}(t,\rho,\Omega)$ is a local operator. 
}
The ``wave function'' for the radial direction can be
rewritten as 
\begin{align}
\psi_{nl} (\rho)=\frac{1}{{\cal N}_{nl}} \sin^l (\rho) \cos^\Delta (\rho)\,
P_n^{l+d/2-1, \Delta-d/2} \left( \cos(2 \rho) \right),
\label{psi}
\end{align}
where $P_n^{\alpha, \beta} (x)$ is the Jacobi polynomial defined by
\begin{align}
P_n^{\alpha, \beta} \left( x \right)=
{(-1)^n \over 2^n n!} (1-x)^{-\alpha} (1+x)^{-\beta}
{d^n \over dx^n}  \left(  (1-x)^{\alpha} (1+x)^{\beta}  (1-x^2)^{n}  \right),
\end{align}
and
the normalization constant ${\cal N}_{nl}$ is given by
\begin{align}
{\cal N}_{nl}=(-1)^n \sqrt{
\frac{ \Gamma(n+l+\frac{d}{2})  \Gamma(n+1+\Delta-\frac{d}{2})   }
{  \Gamma(n+l+\Delta) \Gamma(n+1) }
}.
\end{align}
For later use, we will compute the large $n,l$ limit of ${\cal N}_{nl}$
using the following formula: $\lim_{z \rightarrow \infty} \Gamma(z+a)/\Gamma(z)=z^a$.
For $l \rightarrow \infty$ with a fixed $n$, we have
$(-1)^n {\cal N}_{nl} \rightarrow \sqrt{
\frac{  \Gamma(n+1+\Delta-\frac{d}{2})   }
{  \Gamma(n+1) } }
\frac{1}{l^{\Delta/2-d/4}}$.
For $n \rightarrow \infty$ with a fixed $l$, 
$(-1)^n {\cal N}_{nl} \rightarrow 1 $.
For $n \rightarrow \infty$ with a fixed $l/n$, 
$(-1)^n {\cal N}_{nl} \rightarrow 
\left( 
\frac{n}{n+l}
\right)^{\Delta/2-d/4}
$.
Note that these limiting values are not exponential, but powers of $n,l$.

Then, the bulk local state at $\rho=\rho_0,\Omega=\Omega_0$ is given by
\begin{align}
\ket{\rho_0,\Omega_0,t}
& \equiv 
\hat{\phi}(t,\rho_0,\Omega_0) \ket{0} =
\sum_{n,l,m}  
e^{i t (2n +l+\Delta)}
\psi_{nl} (\rho_0)  Y_{lm} ( \Omega_0)
\hat{a}_{nlm}^\dagger \ket{0}.
\label{bls0}
\end{align}
The other bulk local operators and the bulk local states 
are obtained from these by applying a finite number of space-time derivatives. \footnote{
In the approximation where bulk theory is free, the time derivative does not produce an independent local field by the equations of motion,
except the momentum.
The products and the linear combinations of the local operators
are also local operators, however, we will not consider the products
for simplicity. }

The overlap between $\hat{\phi}(t=0,\rho=\rho_1,\Omega=\Omega_1) \ket{0}$
and $\hat{\pi}(t=0,\rho=\rho_0,\Omega=\Omega_0) \ket{0}$,
where
$\hat{\pi}(t,\rho,\Omega)=\sqrt{-g} g^{tt}
\frac{\partial }{\partial t}\hat{\phi}(t,\rho,\Omega)=
(\frac{\sin \rho}{\cos \rho})^{d-1} \frac{\partial }{\partial t}\hat{\phi}(t,\rho,\Omega)
$ is the momentum which satisfies
$[ \hat{\phi}(t=0,\rho=\rho_1,\Omega=\Omega_1), \hat{\pi}(t=0,\rho=\rho_0,\Omega=\Omega_0)  ]=i \delta(\rho_0-\rho_1) \delta(\Omega_0-\Omega_1)$,
%two different bulk local states 
can be computed from 
\begin{align}
\left( \langle {\rho_1,\Omega_1,t} |\frac{\partial }{\partial t}  \ket{\rho_0,\Omega_0,t} \right)|_{t=0}
=  \sum_{l,m} \left( \sum_{n}  
i(2n +l+ \Delta)
\psi_{nl} (\rho_1) \psi_{nl} (\rho_0) \right) Y_{lm} ( \Omega_1) Y_{lm} ( \Omega_0),
\end{align}
which is (formally) proportional to $i \delta(\rho_0-\rho_1) \delta(\Omega_0-\Omega_1)$ 
because of the orthogonality properties of the Jacobi polynomials and the hyper spherical harmonics.
We will denote 
the bulk local state at $\rho=\rho_0,\Omega=\Omega_0$ on the $t=0$ slice simply as
\begin{align}
\ket{\rho_0,\Omega_0}
& \equiv \sum_{n,l,m}  
\psi_{nl} (\rho_0)  Y_{lm} ( \Omega_0)
\hat{a}_{nlm}^\dagger \ket{0}.
\label{bls}
\end{align}
The overlap of these, $\langle {\rho_1,\Omega_1}   \ket{\rho_0,\Omega_0}$,
may vanishes (formally) except $\rho_0 =\rho_1$ and $\Omega_0 =\Omega_1$.

Of course, these local states (and corresponding local operators) themselves
are not well-defined.
Indeed, they 
have a divergent energy, which means these are ill-defined, and need to be smeared or regularized, for example,
by integrating $\rho, \Omega$ with the Gaussian factors of the center at $\rho_0,\Omega_0$.\footnote{
We can also regularize this by integrating over the time with, for example, the Gaussian factor.
The regularization with the smooth function supported on a spacetime 
region is better for the notion of the local operators in the axiomatic quantum field theory.
We will not use such regularization because it is technically difficult
 for the actual computations and what we will consider below is the low energy effective field theory which is not described by the axiomatic quantum field theory.}
Other way of the regularization are inserting some factors which decays exponentially 
for large $n, l$, for example, $e^{-\hat{H}/\omega_c}$.
Effectively both of these will introduce the cut-off of the contributions for the large $n,l$ in the summations.

Note that if we consider the low energy effective theory with a cut-off of energy $\omega_c$,
there are no local operators smeared over the region smaller than $1/\omega_c$ and
the local operator should be smeared over a region of size, at least, $1/\omega_c$.
%For the radial direction, 
This means that 
there is the largest energy cut-off, around and above which the bulk locality is meaningless
in the effective low energy theory.
In the AdS/CFT correspondence,
the bulk description should be the low energy effective theory.
This is because 
the complete independence assumed in \cite{op} should be 
violated for the finite $N$ CFT and 
the multi particle states, which are related to the
multi trace operators, with a large energy are not independent from the states with lower numbers of particles.
Then, above this energy scale  
the gravity picture (or the bulk locality) is not valid.
This invalidity of the gravity picture
has been known as
the stringy exclusion principle \cite{sep}, at least, for some special examples of $AdS_3/CFT_2$. 
The largest cut-off $\omega_c$  is unknown for general models, however,
it is expected to be $\omega_c  < {\cal O}(N^{2})$
because the degrees of freedom of the CFT is ${\cal O}(N^{2})$.
%the plank scale, 
%i.e. $\omega_c ={\cal O}(N^{2 \frac{1}{d-1}})$.
Therefore, in this paper we consider the bulk locality and the bulk local operators
with the cut-off $\omega_c={\cal O}(N^{2})$ although we will only use 
the fact that a cut-off of the energy is needed and 
we will consider the leading order in $1/\omega_c$ expansion.\footnote{
The gravity picture will be not valid even below this energy scale,
for example, the cut-off will be the plank scale, 
${\cal O}(N^{2 \frac{1}{d-1}})$.
The energy of the local operator will be proportional to the 
cut-off energy and
if the energy of the operator is proportional to the plank scale,
the back reaction to the geometry, which could induces the topology change, should be incorporated.
This back reaction may change the notion of the bulk locality.
Thus, the cut-off is expected to be much less than the plank scale.
}

With the cut-off, %for example,
the overlap of two different bulk local states (\ref{bls}), 
\begin{align}
\langle {\rho_1,\Omega_1} \ket{\rho_0,\Omega_0} 
=  \sum_{l,m} \left( \sum_{n}  \psi_{nl} (\rho_1) \psi_{nl} (\rho_0) \right) Y_{lm} ( \Omega_1) Y_{lm} ( \Omega_0),
\label{ow}
\end{align}
vanishes, up to $1/\omega_c$ corrections, except $\rho_0 =\rho_1$ and $\Omega_0 =\Omega_1$.

We can consider the local state smeared over angular direction $\Omega$, for example by a Gaussian factor,  with length scale $1/l_c$
adding to the energy cut-off $\omega_c$.
This induces the effective cut-off $l_c$ for the summation over $l$ in (\ref{bls}).\footnote{
Because the precise form of the regularization is expected  to be irrelevant,
we will just regard the regularization as the cut-off $l_c$ for the summation over $l$.
}
Let us consider these local states with a regularization satisfies 
\begin{align}
\omega_c \gg l_c,
\end{align}
%\footnote{
%For this, 
%the energy cut-off $\omega_c$ is regarded as the cut-off $\omega_c$ for the summation over $n$ in (\ref{bls}).
%} 
which are important for the discussion below.
For this case, 
we can explicitly see how the overlap between the 
two local states at different space points almost vanishes, as we will see below.
The dominant contributions in the summation over $n$ in (\ref{bls})
are those for $l_c \ll n (\le \omega_c)$
and
the asymptotic behavior of $\psi_{nl} (\rho)$ for large $n$ (with $l$ and $z$ fixed) is
computed, using the asymptotic behavior of Jacobi polynomial \cite{asymptotic}, as 
\begin{align}
\psi_{nl} (\rho) 
%\rightarrow_{n \rightarrow \infty} 
=
\frac{1}{\sqrt{\pi n}} (\tan z)^{\frac{d-1}{2}} 
\cos \left( (2n+l+\Delta) z- \frac{\pi}{2} (\Delta-\frac{d}{2}+\frac12 )\right) +{\cal O}(n^{-3/2}),
\label{cos1}
\end{align}
where 
\begin{align}
z=\pi/2-\rho,
\end{align}
and the boundary is at $z=0$.
In the expression of the overlap of two different bulk local states,
\begin{align}
\langle {\rho_1,\Omega_1} \ket{\rho_0,\Omega_0} 
=  \sum_{l=0}^{l_c}  \left( \sum_{n=0}^{[(\omega_c-l-\Delta)/2]}  \psi_{nl} (\rho_1) \psi_{nl} (\rho_0) \right) \sum_{m} Y_{lm} ( \Omega_1) Y_{lm} ( \Omega_0),
\label{ow1p}
\end{align}
%the contributions from 
the summation over $n$ are almost canceled 
if $\rho_0 \neq \rho_1$, %for the large $n$ 
and if $\rho_0 =\rho_1$ they are almost $l$-independent,
because of the phase factor in (\ref{cos1}) .\footnote{
The $\frac{1}{\sqrt{n}}$ factor in (\ref{cos1}) is canceled if we consider
the commutator between the bulk local field and its momentum. }
Then, for the angular directions, $\sum_{l,m} Y_{lm} ( \Omega_1) Y_{lm} ( \Omega_0) \sim \delta(\Omega_0-\Omega_1)$,
where the large $l$ contributions are almost canceled if $\Omega_0 \neq \Omega_1$
because of phase cancellations.
%Then, we find that the regularized local states satisfies
%$\langle {\rho_1,\Omega_1} \ket{\rho_0,\Omega_0}  \sim \delta(\rho_0-\rho_1) 
%\delta(\Omega_0-\Omega_1)$ approximately 
Here, the resolution of $\rho$ is $1/\omega_c$ and the resolution of the angular direction is $1/l_c$.

Note that these local states with the regularization, which satisfies $\omega_c \gg l_c$,
is squeezed in the radial direction and 
move in radial direction only, thus among the local states, these are a special class of states, of course, because of the condition $\omega_c \gg l_c$.

\subsection{$CFT_d$ local states}

We have considered the bulk local state in $AdS_{d+1}$,
i.e the state given by acting the bulk local field $\phi(t,\rho,\Omega)$
on the vacuum.
On the other hand, the local state in the $CFT_d$, i.e.
${\cal O}_\Delta (t,\Omega) \ket{0} $,
will be different from the bulk local state.
This is possible because the notions of locality are 
different between the finite $N$ $CFT_d$ and the bulk theory.

We can represent the $CFT_d$ local state ${\cal O}_\Delta (t,\Omega) \ket{0} $ as a linear combination of the 
(normalized) energy eigen states,
\begin{align}
c_{nl} s^{\mu_1 \mu_2 \ldots \mu_l}_{(l,m)} P_{\mu_1} P_{\mu_2} \cdots P_{\mu_l}  
(P^2)^n  \hat{{\cal O}}_\Delta \ket{0},
\end{align}
which is identified as 
$\hat{a}_{nlm}^\dagger \ket{0}$.
This rewriting was explicitly done essentially in  \cite{op} by using the hyper spherical Bessel function 
in order to derive
the BDHM \cite{BDHM} extrapolation formula:
\begin{align}
\lim_{\rho \rightarrow \pi/2} 
{ \hat{\phi}(\rho,\Omega)
\over \cos^\Delta(\rho)}
&= 
\sqrt{\pi \over 2}
\sqrt{ \Gamma(\Delta) \over \Gamma(\Delta+1-d/2) \Gamma(d/2)}
{\cal O}_\Delta (\Omega).
\end{align}
If we define 
\begin{align}
\psi_{nl}^{CFT}
\equiv
\sqrt{{2 \over \pi} {\Gamma(\Delta+1-d/2) \Gamma(d/2)  \over \Gamma(\Delta)} }
\lim_{\rho \rightarrow \pi/2} 
{ \psi_{nl} (\rho)
\over \cos^\Delta(\rho)},
\end{align}
where
\begin{align}
\lim_{\rho \rightarrow \pi/2} 
{ \psi_{nl} (\rho)
\over \cos^\Delta(\rho)}
&= 
\sqrt{ \Gamma(n+\Delta+1-d/2) \Gamma(n+l+\Delta)
 \over  \Gamma(n+1) \Gamma(n+l+d/2) }
{1 \over  \Gamma(\Delta+1-d/2)},
\end{align}
the local state in the $CFT_d$ is written as
\begin{align}
{\cal O}_\Delta (\Omega) \ket{0} =
\sum_{n,l,m}
\psi_{nl}^{CFT}
Y_{lm} ( \Omega)
\hat{a}_{nlm}^\dagger \ket{0},
\end{align}
where we have taken $t=0$.
Note that $\psi_{nl}^{CFT} =
\sqrt{{2 \over \pi}   }$ is a constant for $\Delta=d/2$.

For later use, we will compute the large $n,l$ limit of $\psi_{nl}^{CFT}$
up to numerical factors which do not depend on $n,l$.
For $l \rightarrow \infty$ with a fixed $n$, we have 
$\psi_{nl}^{CFT} \rightarrow \sqrt{
\frac{  \Gamma(n+1+\Delta-\frac{d}{2})   }
{  \Gamma(n+1) } }
l^{\Delta/2-d/4}$.
For $n \rightarrow \infty$ with a fixed $l$, we have 
$\psi_{nl}^{CFT} \rightarrow 
{n^{\Delta-d/2}}$.
For $n \rightarrow \infty$ with a fixed $l/n$, 
$\psi_{nl}^{CFT} \rightarrow 
{(n(n+l))^{\Delta/2-d/4}}$.
Note that these asymptotic behaviors are not exponential, but powers of $n,l$.
Note also that the $(2n+l+\Delta)^a\,  \psi_{nl}^{CFT} \rightarrow 
\infty$ for $n \rightarrow \infty$ if $a\geq 1$.

%We have considered the CFT local operator ${\cal O}_\Delta (\Omega)$ 
%and the corresponding CFT local states.
The other local operators in the $CFT_d$ are given from ${\cal O}_\Delta (t,\Omega)$
by acting finite number of spacetime derivatives.
If we are interested in scalar operators only,
the derivatives should be composed by the time derivative and the Laplacian on $S^{d-1}$.
The corresponding local states are given by
\begin{align}
\ket{\Omega,R}_{CFT} & = \left.
R \left(-i \frac{\partial }{\partial t},-\Delta_{S^{d-1}} \right)
\sum_{n,l,m}
 \, e^{i (2n+l+\Delta) t} 
\psi_{nl}^{CFT}
Y_{lm} ( \Omega)
\hat{a}_{nlm}^\dagger \ket{0} \right|_{t=0} 
\\
& =
\sum_{n,l,m}
R(2n+l+\Delta,l(l+d-2)) \, 
\psi_{nl}^{CFT}
Y_{lm} ( \Omega)
\hat{a}_{nlm}^\dagger \ket{0}
\label{lscft}
\end{align}
where $R(x,y)$ is a polynomial of $x,y$.
If we allow an infinite series for $R(x,y)$ it
possibly represent a non-local state.
Below, we will take
$R(x,y)$ as a polynomial of $x$ only and
denote $R(x,y)$ as $R(x)$ for the notational simplicity because 
the $y$-dependence is not relevant in the discussions below.

Because the energy of these states are divergent,
it is important to consider the regularization of these states and the corresponding operators.
Here, we consider the CFT and we can smear local state over some region in time $t$ and space $\Omega$.
For the finiteness of the energy we need to smear, at least, in time
except for the free CFT case.\footnote{
For the free scalar theory, the smearing over space only can give a finite energy state because there is no summation over $n$ for this case.
}
Then, we can introduce, for example, the Gaussian factor $e^{-(2n+l+\Delta)^2/(\omega_c)^2}$,
where $\omega_c \gg 1$, by the time average over a period of $1/\omega_c$, which gives the effective cut-off
$\omega_c$ for $2n+l+\Delta$ 
(and also the effective cut-off $l_c$
for $l$ by the averaging over $\Omega$).

With this regularization, we can not regard 
the state (\ref{lscft}) as a local state in general if the degree of the polynomial $R(x)$ is sufficiently large.
This can be understood as follows.\footnote{
As an example, let us consider 
the free field. The commutation relations at a fixed time,
$[ \frac{\partial^q }{\partial x^q} \phi(x), \pi(x')]= \frac{\partial^q }{\partial x^q}  \delta(x-x')$,
and $[ R(-i \frac{\partial^q }{\partial x^q}) \phi(x), \pi(x')]= R(-i \frac{\partial^q }{\partial x^q})  \delta(x-x')$, where $R(x)=\sum_{q=0}^{q_R} ( i \epsilon x)^{q}/q!$,
vanishe if $x \neq x'$.
For a smeared local field, the commutation relation is
$[ R( -i \frac{\partial^q }{\partial x^q}) \int dy e^{-(\omega_c)^2 (y-x)^2 } \phi(y), \pi(x')]= 
R( -i \frac{\partial^q }{\partial x^q})   e^{-(\omega_c)^2 (x-x')^2 }$.
By taking $\epsilon=x'-x \gg 1/\omega_c$,
this is proportional to $ e^{-(\omega_c \epsilon)^2 }$  for $|\epsilon| \gg q_R/\omega_c$, then, exponentially suppressed,
however, it is almost $1$
for $|\epsilon| \ll q_R/\omega_c$.
}
Let us take $R(x)=\sum_{q=0}^{q_R} ( i \epsilon x)^{q}/q!$ where $\epsilon$ is a real constant
and the $q_R$ is the degree of the polynomial $R(x)$.
Then, if $ | i \epsilon x|^{q_R+1}/(q_R+1) ! \ll 1$, this can not be distinguished with
$e^{i \epsilon x }$, which is the operator generating the translation, $t \rightarrow t+\epsilon$.
Such an operation causes the non-locality of ${\cal O}(\epsilon)$ for time,
and then, the state (\ref{lscft}) at $t=0$ is smeared over the ball shape region with the size $\epsilon$ by the causality, thus it is not a local state.
For a large $q_R$, this condition becomes  $|\epsilon x| \ll q_R$
because $1/q! \sim e^{-q \log q}$.
Because the maximum value of $2n + l+ \Delta$ in  (\ref{lscft}) is ${\cal O}(\omega_c)$,
the state (\ref{lscft}) is supported in a space region of the size $|\epsilon|$ for this $R(x)$ if $|\epsilon| \ll q_R/\omega_c$.\footnote{
Instead of the translation, we can consider $R(x)$ which causes
a smearing over the size $\epsilon$.
For example,
$R(x)=\sum_{q=0}^{q_R/2} ( \epsilon x)^{2q}/q!$, which is a truncation 
of $e^{-( \epsilon x)^{2}}$, is such an operation.
For this $R(x)$ also, 
the state (\ref{lscft}) is supported in a space region of the size $\epsilon$ if $\epsilon \ll q_R/\omega_c$.
}
This implies that, the state (\ref{lscft}) is not necessary regarded as a local state,
at least, if $q_R$ is comparable to $\omega_c$.

Thus, if the cut-off $\omega_c$ is fixed or we consider the low energy effective theory below the energy $\omega_c$, 
the CFT local states (or operators) are effectively local only when $q_R \ll \omega_c$.\footnote{
For the lattice field theory in a box of size $L=1$, with a lattice spacing $1/\omega_c$,
the $q_R$-derivative
$\frac{\partial^{q_R} }{\partial t^{q_R}}$ is an operator
which acts on fields on the $q_R$ sites and induces 
the non-locality of the size $q_R/\omega_c$.
}
This is not inconsistent with the fact that local fields with the arbitrary number of derivatives 
are local fields in a quantum field theory because $\omega_c$ can be taken arbitrary large.
However,
for the AdS/CFT, we will consider a subspace of the Hilbert space of the CFT
which introduces the largest $\omega_c$ because the bulk locality is meaningful only for low energy states. 
(Low energy effective theories for quantum field theories, in general, introduce the largest $\omega_c$.)

Denoting $\omega_c$ as the largest cut-off %(which may be the Plank mass) 
for the bulk picture, 
a CFT state (\ref{lscft}) with the smearing with $\omega'_c$ 
which satisfies $\omega'_c \gg q_R \gg \omega_c$
is effectively local state in the CFT.
However, this state is constructed mostly from the very high energy states with $n \gg \omega_c$.
Such high energy states can not be described in the local bulk picture.

We have seen that the (smeared) CFT local state at a point can effectively becomes the one at a different point by choosing $R(x)$ such that 
it mimics a translation.
This might seem surprising, however, 
for field theories in the Minkowski spacetime, there is the Reeh-Schlieder theorem \cite{RS} which essentially states that the algebra of the local fields on a region in spacetime can generate any state approximately with an arbitrary precision (see, for example, \cite{RS2}).
Thus, the discussions above may be regarded as a (non-rigorous) explicit realization of 
the analogue of the Reeh-Schlieder theorem.

%Below, we will call the CFT states (\ref{lscft})  (or operators) 
%as CFT local states if $q_R \ll \omega_c$, otherwise, CFT states
%because we will only consider CFT operators and states with the regularization by %the smearing over $1/\omega_c$.
Below, we will call the CFT states (\ref{lscft})  (or operators) with $q_R \ll \omega_c$
(more precisely, $q_R ={\cal O} ((\omega_c)^0)$)
as CFT effectively local states  (or operators)
because we will only consider CFT operators and states with the regularization by the smearing over $1/\omega_c$.

Note that the CFT effectively local states only span 
a small subspace of the low energy Hilbert space of the CFT.
However, by the time-evolution, a CFT effectively local state at $t = t_0 \neq 0$ gives 
a CFT state supported in a region at $t=0$, which is not given by effectively local states on $t=0$.
If we consider the CFT effectively local states in whole spacetime (or the spacetime region $\pi \leq t \leq \pi$) and consider 
the time evolution of them to the time slice $t=0$,
then any low energy state is expected to be given by them
because in the low energy theory the effectively local states play the role of the local states.

\section{Bulk Locality in AdS/CFT}

In this section, we will consider the relation between the bulk local states 
and the CFT (effectively) local states. 
This topic is related to
how the bulk locality emerges in the CFT.
The bulk locality is seemingly paradoxical in the CFT point of view,
however, we will see that there may be no paradox.

One of the "paradox" discussed in \cite{ADH} is about the time-slice axiom which,
essentially, says that
there is no non-trivial operator which strictly commutes with 
any local field of CFT. 
However, from the bulk point of view, 
a boundary operator is obtained from the corresponding bulk field 
by taking it to the boundary with the appropriate scaling factor.
This means that there is no bulk local field, strictly speaking.
Indeed, the bulk local field is an approximate notion in the AdS/CFT correspondence
because $N$ should be finite if the $CFT_d$ is well defined as 
a $d$-dimensional field theory.
Thus, this formal "paradox" is not a real problem.\footnote{
Indeed, as we will see below, the boundary operators are the CFT effectively local operators,
which are in a subset of the CFT local operators. 
Thus, the bulk operator can commute with the boundary operators effectively.
}
We will see below how bulk local fields appear in the CFT explicitly.

\subsection{Bulk local states from CFT states}

\subsubsection{Bulk states localized in radial direction}

Here, we will consider bulk states\footnote{
The bulk states and the CFT states are identical, of course.
(In the low energy, it is identical and the CFT may be the definition of the theory above the cut-off scale.)
Here the "bulk state" means that we consider a state, from the bulk point of view, which is obtained
by acting a bulk local field on the vacuum.
}
 which are localized only in the radial direction 
at $\rho=\rho_0$ and 
extended in the angular direction, with $l=l_0, m=m_0$,
%the homogeneous states, i.e. $l=0$ states:
\begin{align}
\ket{\rho_0,l=l_0, ,m=m_0}
& \equiv \sum_{n}  
\psi_{n \, l_0} (\rho_0)  \,
\hat{a}_{n \, l_0 \, m_0}^\dagger \ket{0},
\label{rhol}
\end{align}
with the regularization discussed above. 
These states can be obtained from the bulk local states by an appropriate averaging over angular directions $\Omega$.
Note that for $\rho_0=0$ this is the local state at $\rho=0$ (which implies $l_0=0$ ) and
for $\rho_0 \neq 0$ this is localized in the radial direction, but extended in the angular directions.
For the latter case, if we take $l_0=0$ in particular, the states are uniform in $S^{d-1}$.

Let us consider the time evolution of this state.
At time $t$, this state becomes 
\begin{align}
e^{i \hat{H} t} \ket{\rho_0,l=l_0, ,m=m_0}
& =\sum_{n}  
e^{i (2  n + l_0+\Delta) t}
\psi_{n l_0} (\rho_0)  \,
\hat{a}_{n \, l_0 \, m_0}^\dagger \ket{0}.
\end{align}
Because $\psi_{n l} $ is not singular for finite $n$,
the dominant contributions in the summation are from $n \gg 1$,
and, by using (\ref{cos1}), the phase of each of the contributions becomes
\begin{align}
& e^{i (2  n + l_0+\Delta) t} \cos \left( (2n+ l_0+\Delta) z_0- \frac{\pi}{2} (\Delta-\frac{d}{2}+\frac12 )\right) \CR
= & \frac12 e^{i (2  n + l_0+\Delta) (t+z_0)- \frac{\pi}{2} (\Delta-\frac{d}{2}+\frac12 ) }
+\frac12 e^{i (2  n + l_0+ \Delta) (t-z_0)+ \frac{\pi}{2} (\Delta-\frac{d}{2}+\frac12)} ,
\end{align}
where $z_0=\pi/2-\rho_0$.
%up to the overall phase.
Thus, at time $t$, the states becomes localized at $z=z_0 \pm t$.
This is consistent with the fact that the light-like trajectory 
in the radial direction of the $AdS$ spacetime is $z=z_0 \pm t$
because the localized state in the radial direction has an infinite energy (without the regularization).\footnote{
For $\Delta=d/2$ which saturates the Breitenlohner-Freedman bound \cite{BF}
in the bulk picture, the particle in the bulk travels through the light-like trajectory.
Indeed, for this case,  we can check that $\psi_{n \, 0}(\rho=0) = C  e^{i (2n+\Delta) \pi  /2} \psi_{n \, 0}^{CFT}$,
where $C$ is a $n$-independent constant. This means that the bulk local state at $\rho=0,t=0$
is obtained by the the CFT local state averaged over $S^{d-1}$ at $t=-\pi/2 $ (mod $\pi$).  
}
One might think that this is inconsistent with the HKLL reconstruction
of the bulk local state at the center
for the global AdS \cite{HKLL} 
because
the smearing function used in \cite{HKLL} 
is supported in the region $-\pi/2 \leq t \leq \pi/2$.
However, the smearing function is $K(t)=(\cos t)^{\Delta-d}$ or $K=(\cos t)^{\Delta-d} t$,
then $\int dt K(t)$ is divergent near $t=\pm \pi/2$ for $\Delta \leq d-1$.
Thus, the integration is indeed localized at these points.\footnote{
The divergence is regularized by the smearing of the bulk local operator. For $\Delta > d-1$, some modifications of the discussion in \cite{HKLL} would be needed.}

Let us concentrate on the $l_0=0$ case %mode of $(\ref{rhol})$ 
and 
consider what is the corresponding state as a linear combination of the CFT local states 
(\ref{lscft}) with the regularization.
Because of the rotational symmetry, the corresponding state
should be (\ref{lscft}) with $l=0$
\begin{align}
\ket{l=0,R}_{CFT} \equiv
\sum_{n}
R(2n+\Delta) \, \psi_{n \, 0}^{CFT}
\hat{a}_{n \, 0 \, 0}^\dagger \ket{0},
\label{lscft2p}
\end{align}
with an appropriate $R(x)$.
Thus, the $l=0$ state is uniform in the space $S^{d-1}$ and
can not be described by the CFT state supported 
in any subregion of the space $S^{d-1}$.
This implies that the bulk local state at the center of $AdS$ space are
supported in whole space $S^{d-1}$ and then
%This may imply that 
the version of the subregion duality may not be correct.

For general $l$, we will see how the following CFT state constructed from the CFT states (\ref{lscft})
integrating over the angular direction,
\begin{align}
\ket{l=l_0,m=m_0,R}_{CFT} \equiv
\int d \Omega Y^*_{l_0, m_0}(\Omega) \, 
%R \left(-i \frac{\partial }{\partial t} \right) 
\ket{\Omega,R}_{CFT} 
=
\sum_{n}
R(2n+l_0+\Delta) \, \psi_{n \, l_0}^{CFT}
\hat{a}_{n \, l_0 \, m_0}^\dagger \ket{0},
\label{lscft3}
\end{align}
realizes the bulk state localized in the radial direction  (\ref{rhol}) by choosing $R(x)$ appropriately.
We will also see how the bulk locality in the radial direction is realized in the CFT for this example.

In order to have the bulk state which is localized in the radial direction  (\ref{rhol}) and its $\rho$-derivative,
we need to choose the polynomial $R$ in (\ref{lscft3})
such that the phase factor in the large $n$ is reproduced, i.e.
\begin{align}
R(2n+l_0+\Delta) \, \psi_{n \, l_0}^{CFT}
% \rightarrow
%\frac{1}{\sqrt{\pi n}} (\tan z_0)^{\frac{d-2}{2}} 
\sim 
\exp \left( \pm i \left( (2n+l_0+\Delta) z_0- \frac{\pi}{2} (\Delta-\frac{d}{2}+\frac12 )\right) \right),
\label{p1}
\end{align}
for ${1 \ll n \le \omega_c} $,
where the right hand side is the asymptotic behavior of 
$\psi_{n \, l_0} (\rho_0)$.
This is possible 
because $R$ is an arbitrary polynomial and the range of $n$ considered 
here is finite.
Indeed, if we take
$R(x)=\sum_{q=0}^{q_R} ( i z_0 x)^{q}/q!$ where $z_0  \ll q_R/\omega_c$,
which is effectively a time translation operator,
the state (\ref{lscft3}) reproduces the phase (\ref{p1})
and then the state is localized at $\rho=1/z_0$.
We note that $q_R$ is, at least, comparable with $\omega_c$ which is very large,
thus the CFT state which corresponds to the bulk local state contains 
a large number of derivatives.
Note that with this choice of $R(x)$, 
the CFT local state $\ket{\Omega,R}_{CFT}$ used in (\ref{lscft3}) is not a CFT effectively local state.

Thus, the CFT state (\ref{lscft3}) with the appropriate choice of $R(x)$ describes
the bulk local states (\ref{rhol}). %with $l=0$
%although the CFT primary operator corresponds to the operator on the boundary of the bulk.
This is possible because of the large degrees of freedom 
of the large $N$ gauge theory, of course. 
The different states at different $\rho$, which are orthogonal each other, are embedded in the 
${\cal O} (N^2)$ internal degrees of freedom of the CFT,
which essentially correspond to the label $n$ and
the $n$-dependence of the state is changed by the time derivatives.
A linear combination of the CFT local states with the large number of time derivatives corresponds to 
the bulk state away from the boundary.

\subsubsection{Examples of bulk local states from CFT states}

We have seen that the bulk local state at the center corresponds 
to the CFT state which is supported on whole space $S^{d-1}$ uniformly.
This implies that a generic bulk local state is also 
corresponds 
to a CFT state which is supported on whole space,
because of the continuity.

On the other hand, some particular bulk local states
can correspond to CFT states which is supported on a subregion in the space
as we will see below.\footnote{
%We note that because of the regularization, the bulk local operator at a point is not unique. More precisely, 
The bulk local state constructed here 
is a local state with the resolution $1/l_c$, but not a local state with 
the resolution $1/\omega_c$. 
Thus, precisely speaking, the bulk local state constructed here 
is not an effectively local state. 
}
Let us consider the CFT state (\ref{lscft}) 
at $\Omega=\Omega_0$
with the regularization,
\begin{align}
\ket{\Omega=\Omega_0,R}_{CFT} & = 
\sum_{n,l,m}^{2n +l+ \Delta <\omega_c, \, l<l_c}
R(2n+l+\Delta) \, 
\psi_{nl}^{CFT}
Y_{lm} ( \Omega_0)
\hat{a}_{nlm}^\dagger \ket{0}.
\label{lscft0}
\end{align}
This state is a CFT local state smeared over the time with
the length scale $1/\omega_c$ and over the space with the length scale $1/l_c$.
Here, we take the parameters of the regularizations as $1 \ll l_c \ll \omega_c$.
($1/\omega_c$ can be the smallest one, which may be related to the plank length and
$1/l_c$ will be an arbitrary very small length scale, but much larger than $1/\omega_c$.)

If this CFT state 
corresponds to
the bulk local state (\ref{bls}) at $\rho=\rho_0,\Omega=\Omega_0$,
the phase factor should be reproduced.
For this, we need to take 
\begin{align}
R(x)=\sum_{q=0}^{q_c} ( i z_0 x)^{q}/q!, 
\end{align}
where $z_0  \ll q_c/\omega_c$
as in (\ref{p1}).
Then, if we neglect $1/l_c$ suppressed terms,
the overlap between the (normalized) CFT state (\ref{lscft0})
and the bulk local state (\ref{bls}) at $\rho=\rho_1,\Omega=\Omega_1$ with the regularization by $\omega_c$,
\begin{align}
\langle \rho_1,\Omega_1 \ket{\Omega=\Omega_0,R}_{CFT} =
\sum_{n,l,m}^{2n +l+ \Delta <\omega_c, \, l<l_c}
R(2n+l+\Delta) \, \psi_{nl}^{CFT}
\psi_{nl}(\rho_0)
Y^*_{lm} ( \Omega_1)
Y_{lm} ( \Omega_0),
\label{ow1}
\end{align}
divided by the normalization factors of the two states 
vanishes\footnote{
For $\omega_c \sim l_c$, the asymptotic behavior of $\psi_{n l}$ is 
complicated %and the phase factor depends on $n$ and $l$ 
\cite{asymptotic2}, thus we can not construct a generic bulk local state from
a CFT local state..
} except $\rho_0=\rho_1$ up to $1/\omega_c$ and $\Omega_0=\Omega_1$ up to $1/l_c$ because 
$\sum_{l,m} Y^*_{lm} ( \Omega_0) Y_{lm} ( \Omega_1) \sim \delta(\Omega_0-\Omega_1)$.
Thus, this CFT state (with the finite number of the space-time derivatives)
corresponds to a bulk local state at
$\rho=\rho_0=1/z_0$ and $\Omega=\Omega_0$ smeared over the small space time region.

Note that the condition $1 \ll l_c \ll \omega_c$ implies that
the corresponding bulk local state %for $z_0=0$ 
may have 
the momentum along the radial direction only.\footnote{
 $1 \ll l_c \ll \omega_c$ implies  $1 \ll l_c \ll n_c$ where 
$n_c$ is the largest $n$ in the sum.
Note also that the momentum is incoming in the radial direction, but,
by changing the $R(x)$ to $R(-x)$, it becomes outward.
}
This interpretation is consistent with the fact that the CFT state 
is the bulk local state at $z=z_0$ which is moving into the center
along the radial direction from the boundary at $t=-z_0$.
The bulk local state at the center with $l=0$ can be obtained by the superposition
of such CFT states with $z_0=\pi/2$ averaging over the whole space. 

Note also that this CFT state, which is approximated as
\begin{align}
\ket{\Omega=\Omega_0,R}_{CFT} & \approx 
\sum_{n,l,m}^{2n +l+ \Delta <\omega_c, \, l<l_c}
e^{i z_0 (2n+l+\Delta)}  \, 
\psi_{nl}^{CFT}
Y_{lm} ( \Omega_0)
\hat{a}_{nlm}^\dagger \ket{0},
\label{lscft2}
\end{align}
is not a CFT effectively  local state 
because of the large number of the derivatives in $R$ which 
is regarded as a time translation operator by $z_0$.
Thus, this CFT state is supported on the approximately sphere shape ($S^{d-2}$) region with radius $z_0$
in the space $S^{d-1}$,
as we will explicitly see below.
The overlap between this state and a CFT local state at $\Omega=\Omega_1$ 
associated with a polynomial $\tilde{R}(x)$ is
\begin{align}
& {}_{CFT} \langle \Omega=\Omega_1,\tilde{R} \ket{\Omega=\Omega_0,R}_{CFT}  \CR
 \approx &
\sum_{n,l,m}^{2n +l+ \Delta <\omega_c, \, l<l_c}
e^{i z_0 (2n+l+\Delta)}  \tilde{R}^* (2n+l+\Delta) \, 
(\psi_{nl}^{CFT})^2 \,
Y_{lm}^* ( \Omega_1)
Y_{lm} ( \Omega_0) \CR
 =&
\sum_{l,m}^{ l<l_c}
\left(
\sum_{n}^{2n +l+ \Delta <\omega_c}
e^{i z_0 (2n+\Delta)}  \tilde{R}^* (2n+l+\Delta) \, 
(\psi_{nl}^{CFT})^2 
\right)
\,
Y_{lm}^* ( \Omega_1) \,
e^{i z_0 l} \,
Y_{lm} ( \Omega_0).
\label{owcc}
\end{align}
We take $\ket{\Omega=\Omega_1,\tilde{R}}_{CFT}$
as a CFT effectively local state at $\Omega=\Omega_1$, i.e .
the degree of the polynomial $\tilde{R} (x)$ is much less than $\omega_c$.
Then, the summation of $n$ in the last line of (\ref{owcc})
is almost canceled by the phase factor and exponentially suppressed,
except the terms for $n < {\cal O} (1/z_0)$.\footnote{
The overlap (\ref{owcc}) is small,
if we normalize $\ket{\Omega=\Omega_1,\tilde{R}}_{CFT}$ by choosing the overall factor of $\tilde{R} (x)$ appropriately,
because of the cancellation.
However, the number of the independent CFT effectively local states
are large due to the choice of $\tilde{R} (x)$.
}  
These remaining terms are not exponentially large for large $l$
because $(\psi_{nl}^{CFT})^2 \sim l^{\Delta-d/2}$.
We also see that 
$\sum_{l,m}^{ l<l_c} Y_{lm}^* ( \Omega_1) \,
e^{i z_0 l} \,
Y_{lm} ( \Omega_0)$ almost vanish except 
$\Omega_1 \in  S^{d-2}_{z_0}(\Omega_0)$
up to $1/l_c$ corrections
where $ S^{d-2}_{z_0}(\Omega_0)$
is the sphere shape ($S^{d-2}$) region whose center is at $\Omega=\Omega_0$ with radius $z_0$
in the space $S^{d-1}$.
To see this, let us consider the free CFT, i.e. $\Delta=d/2-1$, although it is not holographic CFT.
For this case, there are no summation over $n$ and the overlap
(\ref{owcc}) becomes 
\begin{align}
& {}_{CFT} \langle \Omega=\Omega_1,\tilde{R} \ket{\Omega=\Omega_0,R}_{CFT}  \CR
 =&
\sum_{l,m}^{ l<l_c}
\left(
e^{i z_0 \Delta}  \tilde{R}^* (l+d/2-1) \, 
\frac{2d-1}{\pi(l+d/2-1)}
\right)
\,
Y_{lm}^* ( \Omega_1) \,
e^{i z_0 l} \,
Y_{lm} ( \Omega_0),
\label{owccp}
\end{align}
which should almost vanish except the case that  
$\Omega_0$ can be reached at $t=z_0$ by a light ray, which is described by the free CFT, emanating from 
$\Omega=\Omega_1$ at $t=0$.
This means that $\sum_{l,m}^{ l<l_c} Y_{lm}^* ( \Omega_1) \,
e^{i z_0 l} \,
Y_{lm} ( \Omega_0)$ has the property stated above.
Therefore, the CFT state  (\ref{lscft2})
is supported on the sphere region from the CFT point of view
although it is the bulk local state form the bulk point of view.\footnote{
For $z_0=\pi$, the CFT state is a CFT effectively local state at the opposite point ($\bar{\Omega}_0$)
to $\Omega=\Omega_0$ in $S^{d-1}$ as we can see it explicitly by using $e^{i \pi l} \,Y_{lm} ( \Omega_0)=Y_{lm} ( \bar{\Omega}_0)$. 
}

In order to understand how the bulk local state appears from the CFT clearer,
let us consider $d=2$ case.
For this case, the CFT state at $t=0$ is supported on two points $S^0$ which is a time(-reversed) evolution of 
the CFT effectively local state at $\Omega=\Omega_1, t=z_0$.
Then, one might think that 
the ``particles'' at the two points are entangled and regarded as an EPR pair.
%because no entanglement means that we can regard the two ``particles'' 
%as two independent CFT effectively local states at the two points, which correspond
%to a bulk state near the boundary.
However, 
These two ``particles'' are just a superposition although the EPR pair is a two particle state.
Thus, the bulk local state is a non-local state in the CFT and
this non-locality is related to the ``basis'' change of the states, which dose not keep the locality.
This also means that the state is an entangled state.

\subsection{CFT states supported in a space region from bulk local states}

As we have seen, the bulk local state is, in general, reconstructed from 
the states in the CFT supported in the whole space $S^{d-1}$.
(This fact was also known in the HKLL reconstruction in the global coordinate \cite{HKLL}.)
However, the CFT effectively local states should be supported 
in some region in the bulk, in the bulk point of view, because of the causality.
For example, the CFT local operator at the north pole of $S^{d-1}$ and 
the one at the south pole should commute each other 
if the difference of time is shorter than $\pi$ by the causality in the CFT.\footnote{
Our smearing of the fields is not restricted on a region, but  
it is like a Gaussian. Thus, the commutator vanishes with some terms suppressed by the regularization parameters.
}
This is impossible if the corresponding bulk states are extending in the bulk
sufficiently because the propagation in the bulk can connect two boundary regions faster than 
what the CFT requires, as shown in the Appendix \ref{lr}. 
Note that there is only one free scalar field in the bulk corresponding to the large $N$ CFT fields,
i.e. the degrees of freedom of the bulk theory is one (or ${\cal O}(1)$), not ${\cal O}(N^2)$.
Thus, if the propagation in the bulk connect two spacetime points,
two generic local operators at these points will have a non-zero commutator.

In this section, 
we will study bulk interpretations of the CFT effectively local state and
the CFT states which are supported in a region and 
see how these states extends in the bulk.\footnote{
In \cite{Takayanagi}, similar discussions have been done for the 
Euclidian time evolution.
}

\subsubsection{CFT effectively local states}

Here, we will explicitly consider bulk interpretation of the CFT local state $\ket{\Omega_0,R}_{CFT} $ at $\Omega=\Omega_0$ 
with the regularization by the smearing for the time direction 
and the space ($S^{d-1}$) direction with the resolutions given by $1/\omega_c$ and $1/l_c$, respectively:
\begin{align}
\ket{\Omega_0,R}_{CFT} =
\sum_{n,l,m}^{2n +l+ \Delta <\omega_c, \, l<l_c}
R(2n+l+\Delta) \, \psi_{nl}^{CFT}
Y_{lm} ( \Omega_0)
\hat{a}_{nlm}^\dagger \ket{0}.
\end{align}
This state is a CFT effectively local state by requiring that the degree $q_R$ of the polynomial $R(x)$ is small ($\omega_c \gg q_R$) and 
the derivatives do not change the large $n$ behavior of the phase factor.
Here, we concentrate on the case that the smearing for the space direction is much larger than the one for the time direction,
which means $1 \ll l_c \ll \omega_c$.\footnote{
We can also consider the case with $l_c \sim \omega_c$ or the case with the smearing for the time direction only 
by using the results in \cite{asymptotic2}. 
Here, we consider the case with $1 \ll l_c \ll \omega_c$ only because it is simpler and 
sufficient for showing where in the bulk the CFT state extends.
}
%as shown in XXXX,
The overlap between this state
and the bulk local state (\ref{bls}) at $\rho=\rho_1,\Omega=\Omega_1$ is given by (\ref{ow1})
and, for this case, 
the phase factor in the sum for the $n$ is
\begin{align}
R(2n+l+\Delta) \, \psi_{n \, l}^{CFT} \psi_{nl} (\rho_1)
% \rightarrow
%\frac{1}{\sqrt{\pi n}} (\tan z_0)^{\frac{d-2}{2}} 
\sim 
\cos \left( (2n+l+\Delta) z_1- \frac{\pi}{2} (\Delta-\frac{d}{2}+\frac12 )\right) ,
\label{uvir}
\end{align}
for ${l \ll n \le \omega_c} $.
Thus, the overlap 
almost vanishes except $z_1(=\pi/2-\rho_1)=0$ up to $1/\omega_c$ and $\Omega_0=\Omega_1$ up to $1/l_c$.
This means that the CFT effectively local states corresponds to the bulk local states near the point on the boundary. 
In other words,
the CFT primary states (or fields) with a sufficiently small number of derivatives (and a low conformal dimension)
live on the boundary of the bulk theory.
This statement is a refinement of the statement ``CFT lives on the boundary of AdS'' in AdS/CFT correspondence
although the CFT corresponds to the whole bulk theory.
%Note that the notion of the locality in the bulk depends on the regularization $\omega_c$.

If we change the high energy cut-off $\omega_c$ lower,
the CFT effectively local state with this cut-off 
extends in the radial direction in the bulk for $z < {\cal O}(1/\omega_c)$. 
This may be
a kind of a realization of the UV/IR relation.

\subsubsection{CFT states supported in a region}

Let us consider CFT states effectively supported in a space ($S^{d-1}$) region $A$ and 
we call the subspace of the Hilbert space spanned by them as ${\cal H}^{eff}_{A}$.
Because we consider only the low energy states, the states are 
really supported in spacetime region which is given by extending the space region to time direction with 
the width $1/\omega_c$. 
Note that the CFT (formal) local states, which are obtained by acting the CFT local operators on the vacuum,
are not included in ${\cal H}^{eff}_{A}$ generically because they are effectively non-local even if they are smeared over the region $A$.
Note also that the linear combinations of the CFT effectively local states in the space region $A$
are in ${\cal H}^{eff}_{A}$, however, there exist other states in ${\cal H}^{eff}_{A}$, corresponding to
the states inside the bulk,
as we have seen.
Indeed, the CFT state (\ref{lscft2}) is not a CFT effectively local state, but is 
supported in $S^{d-2}$ sphere with radius $z_0$ around $\Omega=\Omega_0$.
Thus, this state is in ${\cal H}^{eff}_{A}$ if we choose $z_0$ appropriately
small and $\Omega_0 \in A$.

Note that 
the CFT states which are 
supported in the $S^{d-2}$ sphere around $\Omega=\Omega_0$ with radius $z_0$, or the ball including the inside of this sphere,
can not correspond to a bulk state including a bulk local state at $z>z_0, \Omega=\Omega_0$
because of the causality discussed above and the results in the Appendix \ref{lr}.
The CFT state (\ref{lscft2}) which corresponds to a bulk local state at $z=z_0, \Omega=\Omega_0$,
has the maximum value for $z$ under this causality bound for the states in ${\cal H}^{eff}_{A}$.

Let us consider 
the bulk states correspond to 
the CFT states supported in this ball shaped region $A$ (at $t=0$).
These CFT states will be generated by considering the all CFT effectively local states 
in the causal diamond of the space region $A$ (the boundary domain of dependence of $A$) and 
the corresponding CFT states at $t=0$ by the time evolution.\footnote{
CFT states which enter in the causal diamond by the time evolution
are also supported in the region $A$, however, the corresponding CFT states at $t=0$
are represented by the CFT states in the causal diamond also.
}

We expect that the CFT effectively local states at a point $p$ in the causal diamond can represent any bulk local state,
which can move various directions,
at the point $p$ which are on the boundary of the bulk spacetime.
Thus, the region $a$ in $t=0$ slice in the bulk which can be reached by the bulk local states in the causal diamond
by the time evolution 
is the bulk region corresponding to the region $A$
and we can see that this region $a$ coincides with the causal wedge of $A$.
This means that 
the CFT states supported in the region $A$
are given by the bulk states supported in the causal wedge of $A$.

This is similar to the version of the subregion duality, however, there are no inverse of this, 
thus this is not like a duality.
As we have seen, there are some
bulk states supported in the causal wedge of $A$
which can not be given by the CFT states supported in the region $A$.
Indeed, the CFT state (\ref{lscft2}) corresponds to a bulk local state at $z=z_0, \Omega=\Omega_0$
is moving along the radial direction and it may not be possible to 
correspond to a bulk local state at the point moving in different direction.\footnote{
One might think that the superposition of different local states can 
move another direction. However, such superposition is not a bound state and
move different directions separately.
In particular, by an averaging the CFT states over a time period of $1/l_c$ near $t=z_0$.
we obtain a CFT state consist with bulk local states moving in all directions
and the corresponding bulk local state at $z=z_0, t=0$, which is a small part of it, is moving in radial direction.
}
One might think that by changing the phases in  (\ref{lscft2}) one could build a wavepacket that describes an excitation moving in any direction and able to represent any state in the causal wedge of A. But, if change the phases by hand (or further smearing over the radial direction), then the state is no longer supported in CFT region A, generically. Thus, it is impossible to construct such wave packet from the CFT states in A. 

Thus, a bulk local state is decomposed to the bulk local states moving in particular directions, each of which are represented by the CFT state (\ref{lscft2}) or the rotation of it.
some of these rotated CFT states (\ref{lscft2}) are supported in the region  $A$ and
the others are not supported in $A$.
The bulk local state constructed from the CFT states supported in a region $A$
is a sum of such CFT states, but only for the ones supported in the region $A$.
This also implies that
bulk local states (or operators) at a same bulk point constructed from
CFT states supported in different regions are different even in the low energy (gravity) theory.\footnote{
The bulk local state 
moving in some direction is not a local state with the resolution $1/\omega_c$.
Thus, precisely speaking, the bulk local state can not be constructed from
the CFT states supported in the subregion.
}
Note that in the quantum error correction code proposal \cite{ADH},
such two bulk local states were assumed to be same in the code subspace, which may corresponds to 
the subspace of the low energy gravity theory.
Thus, in our example, the quantum error correction code proposal may not be realized.

\subsubsection{Entanglement wedge or causal wedge?}

As we have seen,
for the connected region $A$,  
the CFT states supported in the region $A$
are given by the bulk states supported in the causal wedge of $A$.
The causal wedge of $A$ and entanglement wedge of $A$ are same for this case.
On the other hand, these two wedges can be different for a region $A$ which is a sum of disconnected regions $A_1$ and $A_2$,
i.e. $A=A_1 \cup A_2 $ and $A_1 \cap A_2=\varnothing$.
For this case, there are bulk states which are not contained in 
bulk states supported in $a_1 \cup a_2$
where $a_i$ is the causal wedge of $A_i$.
Let us consider, for example, the case where $A_i$ are very small regions and are regarded as points.
Then, their is a spacetime point $p$ of CFT (or the boundary of the bulk) which is a midpoint of $A_1$ and $A_2$ in space
and at the (past) time such that the point $p$ is connected to $A_1$ and $A_2$ by light like trajectories.
Then the  time evolution to $t=0$ of a CFT effectively local state at $p$ may be 
supported in $A=A_1 \cup A_2$ in the CFT point of view.
In the bulk viewpoint, a bulk state at $p$ can move toward the center (i.e. in radial direction) or toward $A_i$ 
in space.
Moreover, it can move in any direction between these
and will reach different bulk point at $t=0$.
Then, for $d=2$ case, the bulk state at $t=0$ will be supported on a minimal surface (curve) in the bulk connecting $A_1$ and $A_2$
because the minimal surface is the boundary of
the causal wedge of the segment which is given by the geodesics between $A_1$ and $A_2$.
For general $d$, 
this statement will be correct although the minimal surface is degenerated.
Thus, for this $A$, the CFT states supported in the region $A$
correspond to the bulk states supported in 
the sum of the causal wedge of $A$ and this region given by the minimal surface.

Now, let us consider the case where $A_i$ are not small regions.
For this case, 
the bulk states at $t=0$ will be supported in the region which 
is the sum of the (degenerate) minimal surface between the all possible
points in $A_1$ and $A_2$.
This region is expected to be 
the bulk region whose boundaries are $A_i$ and the minimal surface connecting 
these in the space at $t=0$.
Note that
the entanglement wedge is the region 
surrounded by the 
Ryu-Takayanagi surface \cite{RT} for $A$, which 
is this region or the causal wedge of $A$, 
depending on the area of the minimal surfaces.
We will call the sum of regions surrounded by any possible minimal surface whose boundaries are in the  
boundary of $A$ as minimal surface wedge of $A$.
Then, the minimal surface wedge may be the bulk region corresponding to
 $A$, i.e. any CFT state supported in $A$ may be expressed by a bulk state supported in the minimal surface wedge of $A$.\footnote{ 
We expect that this is true for any region $A$.}

Note that it is known that the causal wedge is always inside the entanglement wedge for our case.
However, the bulk local state at a space point $p$ in the causal wedge 
constructed from  $A_1$ (or $A_2$) only
is different from the above bulk local state at the same point $p$
constructed from both $A_1$ and $A_2$.
According to \cite{RT}, the entanglement entropy is proportional to the area of the 
Ryu-Takayanagi surface.
It is interesting to understand this Ryu-Takayanagi formula explicitly in the view point of this paper.

\subsection{Comments on subregion duality and quantum error correction}

As we have seen, 
the version of the subregion duality is not valid in our analysis.
A bulk local state corresponds to a CFT state supported in whole space,
although a CFT state supported in a region $A$ will correspond
to a bulk state supported in the minimal surface wedge of $A$.
This is possible because a bulk state constructed from CFT states in a region $A$
is different from a bulk state at the same point constructed from CFT states in a region $A'$ which is different from $A$.

Our analysis is essentially same as the global AdS reconstruction of HKLL.
Instead of it, the AdS-Rindler reconstruction was also discussed in \cite{HKLL}.
This reconstruction was used to argue that the quantum error correction is relevant for quantum gravity in \cite{ADH}.
However, the AdS-Rindler wedge only cover a part of the AdS spacetime
and we need continuum spectrum for the mode expansion for construction of the smearing function for it.
The spectrum of the CFT is discrete, where the Hamiltonian is the dilatation.\footnote{
In order to define the bulk operator, we need to choose the Hamiltonian
because the bulk local operator will be defined in the low energy theory
and the low energy states on which the bulk local operators act are defined 
by the Hamiltonian.}
Thus we need infinite energy modes of the CFT for the AdS-Rindler reconstruction\footnote{
The continuum spectrum is due to the boundary condition for the Rindler wedge
although the CFT states satisfies the ``periodic boundary condition'' for $S^{d-1}$.
}
and, the bulk state from the AdS-Rindler reconstruction is unclear 
in the CFT states, in particular for the low energy theory.

If we forget about the CFT states and Hamiltonian, based on which we have discussed 
low energy theory, the AdS-Rindler Hamiltonian will be no problem and 
there will be duality between the subregions of the bulk and CFT, like the AdS-Poincare reconstruction.
However, if we consider the two different regions, $A$ and $A'$, 
the AdS-Rindler reconstructions depend on the different Hamiltonians on the two regions
and it is unclear to how to compare the two bulk local states constructed in the different regions
in the low energy theories, in particular the two bulk local states will not be equivalent in the low energy theory.
Thus, it is not justified to use the AdS-Rindler reconstruction 
for the discussion on the bulk local operators for such cases.

In our analysis, a CFT state supported in a ball shaped region $A$ 
will correspond to a bulk state supported in the causal wedge of $A$,
which also appears in the AdS-Rindler reconstruction, 
however, 
two bulk local states at a same spacetime point constructed in different regions are different
even in the low energy theory. Thus, the theory does not have the structure of 
quantum error correction codes discussed in \cite{ADH}.

The bulk local operator of the AdS-Rindler reconstruction will reproduce 
the bulk correlation function. However, 
the correlation function.with the same bulk local operator insertions  will be different 
each other for the AdS-Rindler reconstruction and the global reconstruction.
This is because that AdS-Rindler reconstruction
used the eigen modes with the  (ingoing) boundary condition in the bulk, then
it reproduces the bulk correlation function with the boundary condition,
which is absent for the global reconstruction.
Thus, the bulk local operator of the AdS-Rindler reconstruction 
should be different from the bulk local operator of the global reconstruction even in the low energy.

%
%Why localization in radial direction appears?
%Why causality in radial direction appears?
%Conformal symmetry will not useful because 
%non-conformal theory will be causal and local
%where only the near-boundary region is related 
%to conformal symmetry.
%
%n 1/N expansion (inclusion of the interaction in AdS), 
%locality and causality will hold.
%Which property of CFT guarantees it?
%Anyway, what is a definition of 1/N expansion in CFT?
%
%What state in CFT does correspond to classical geometry in asymptotic AdS? Why the locality and causality appear?
%
%By the lattice like construction diffeo will always
%appear. but, graviton will not usually. Why?
% 
%Black hole formation?
%
%Coherent states of $a^\dagger$ as semi-classical geometry?
%
%SUSY to KK states
%
%4d N=4 from SQM then, large N limit gives QG?
%
%UV-IR

\section*{Acknowledgments}

S.T. would like to thank  
Kanato Goto, Lento Nagano, Yu Nakayama, Yausnori Nomura, Sotaro Sugishita, Yuki Suzuki, Tadashi Takayanagi, Kotaro Tamaoka and Koji Umemoto
for useful discussions.
S.T. would like to thank the Yukawa Institute for Theoretical Physics at Kyoto University.
Discussions during the workshop YITP-T-19-03 "Quantum Information and String Theory 2019"
were useful to complete this work.
This work was supported by JSPS KAKENHI Grant Number 17K05414.

\vspace{1cm}

%\noindent
%{\bf Note added}: 

\appendix

\section{On the causality on AdS/CFT}

\label{lr}

The metric of $AdS_{d-1}$ spacetime is given by 
\begin{align}
d s^2_{AdS} = \frac{1}{\sin^2 (z)} 
\left( 
-dt^2 +d z^2+ \cos^2 (z)  d \Omega_{d-1}^2 
\right).
\end{align}
where $0 \leq z \leq \pi/2$.

Let us consider a 
light-like trajectory along the radial direction from a bulk point near 
the north pole ($z=\epsilon, \theta=0$) at $t=0$ to the south pole ($z=0, \theta=\pi$) where $\theta$ is an angle variable of $S^{d-1}$ satisfying $0 \leq \theta \leq \pi$. 
For this trajectory, $dt=d z$ and  
it will be at the south pole at $t= \pi-\epsilon$.
Instead of this, let us consider 
a light-like trajectory along the angular direction from a bulk point on the boundary near 
the north pole ($z=0, \theta=\epsilon$) at $t=0$ to the south pole ($z=0, \theta=\pi$)
with $d z=0$
satisfies $dt=d \theta$, thus it will also be at the south pole at $t= \pi-\epsilon$.
Note that this light-like trajectory is regarded as the one in the $CFT_{d}$
on $S^{d-1}$ as a space.

Thus, the CFT state supported in a ball shaped region of size $\epsilon$ centered at the north pole of  $S^{d-1}$
can not correspond to the bulk local state at $z>\epsilon, \theta=0$
because of the causality of the CFT.

The following is not relevant for the discussions in this paper, but,
we can consider 
a light-like trajectory from a bulk point near 
the north pole ($z=\epsilon, \theta=0$) at $t=0$ to a point near the south pole ($z=\epsilon', \theta=\pi$) not along the radial direction.
It satisfies
$dt=d \theta \sqrt{{d z^2 \over d\theta^2}+ \cos^2 (z(\theta)) } 
> d \theta \cos(z(\theta))$.
Thus it will be at the point after $t= \pi-\pi (\epsilon')^2/2$,
for $\epsilon \ge \epsilon'$,
where we assumed $\epsilon \ll 1$.

\end{document}